# An opto-electro-mechanical system based on evanescently-coupled optical microbottle and electromechanical resonator

Motoki Asano[1], Ryuichi Ohta[1], Takashi Yamamoto[2], Hajime Okamoto[1], and Hiroshi Yamaguchi[1]

[1]NTT Basic Research Laboratories, NTT Corporation, 3-1 Morinosato Wakamiya, Atsugi-shi, Kanagawa 243-0198, Japan
[2]Department of Materials Engineering Science, Graduate School of Engineering Science, Osaka University, Toyonaka, Osaka 560-8531, Japan

Evanescent coupling between a high-Q silica optical microbottle and a GaAs electromechanical resonator is demonstrated. This coupling offers an opto-electro-mechanical system which possesses both cavity-enhanced optical sensitivity and electrical controllability of the mechanical motion. Cooling and heating of the mechanical mode are demonstrated based on optomechanical detection via the radiation pressure and electromechanical feedback via the piezoelectric effect. This evanescent approach allows for individual design of optical, mechanical, and electrical systems, which could lead to highly-sensitive and functionalized opto-electro-mechanical systems.

Controlling mechanical motion has been actively studied in various optomechanical[1] and electromechanical systems[2]. Hybridizing these systems, namely constructing an opto-electro-mechanical system, enables us to integrate optical and electrical controllability of mechanical motion for applications such as coherent conversion between microwave and optical photons[3]. Moreover, it is possible to implement cavity optomechanics assisted by electrically excited mechanical nonlinearity[4-7], such as mechanical squeezing by Duffing-type mechanical oscillator[8] and squeezing-enhanced optomechanical coupling[9]. Opto-electro-mechanical systems have been developed with state-of-the-art optomechanical devices, such as photonic crystals[10-14] and whispering-gallery-mode (WGM) resonators[15], by installing electrodes to the devices to excite mechanical motion via capacitive or piezoelectric electromechanical conversion. These systems allow for strong optomechanical coupling thanks to a large overlap between optical and mechanical modes[16,17]. However, they lack the flexibility in design and material of the electromechanical system because they are based on the optimal optomechanical architecture.

An alternative approach to construct an opto-electro-mechanical system would be to form it on the optimal electromechanical devices. One way to achieve this, which we demonstrate here, is to evanescently couple an optical resonator (cavity) to an electromechanical resonator. Such an approach enables us to develop an opto-electro-mechanical system which allows for individual design of optical, mechanical, and electrical systems. In this letter, we report on the development of an opto-electro-mechanical system in which a movable high-Q silica optical microbottle[18-23] is evanescently coupled to a piezoelectric GaAs electromechanical resonator[24-26]. The electromechanical resonator is doubly clamped and Au electrodes are put on to allow electric access to the mechanical resonator via the piezoelectric effect. Approaching the high-Q optical microbottle resonator to the electromechanical resonator using a nano-positioner allows for both optical detection and electrical manipulation of the mechanical motion in the GaAs resonator.

Figure 1(a) is an illustration of the silica optical microbottle resonator with a tapered optical fiber which was used to couple the light into WGMs. The evanescent field of WGMs plays a central role in inducing the optomechanical coupling between the optical microbottle resonator and piezoelectric mechanical resonator. The silica microbottle resonator was fabricated by the heat-and-pull technique from standard silica optical fiber (clad diameter: 80 μm). The maximum diameter, the diameter of the necks, and the distance between the two necks were 80 μm, 58 μm, and 0.8 mm, respectively (see Fig. 1(b)). Figure 1(c) is an illustration of the GaAs electromechanical resonator which forms a doubly clamped beam structure. The GaAs electromechanical resonator (150-μm-long and 20-μm-wide) was fabricated from a 600-nm-thick AlGaAs/GaAs modulation-doped heterostructure (95-nm-thick Si-doped $Al_{0.3}Ga_{0.7}As$ on 400-nm-thick GaAs) on a 3-μm-thick $Al_{0.65}Ga_{0.35}As$ sacrificial layer (see Fig.1 (d)). The piezoelectric AlGaAs layer was sandwiched between the Schottky electrode and conductive layer contacted to the ohmic electrode. By applying voltage between them ($x$-direction), piezoelectric stress is generated along the beam direction ($y$-direction) in the AlGaAs layer. Since the beam is doubly clamped, this stress in the upper (AlGaAs) layer of the beam leads to the bending moment. Thus, with ac voltage, the flexural mechanical motion ($x$-direction) can be excited electrically. Note that the electrical conduction between the two mechanical supports through a two-dimensional electron system (2DES) is isolated by a shallow mesa structure on the beam.

Figure 1(e) is a schematic image of the experimental setup to characterize the opto-electro-mechanical properties. Optomechanical coupling takes place thanks to the mode overlap between the mechanical motion and the evanescent field. The gap $d_{\text{OM}}$ between the microbottle resonator and electromechanical beam was adjusted by using a three-axis nano-positioner in a vacuum of $10^{-5}$ Pa (see Fig.1 (f)). Before starting the measurement, we optimized the horizontal position of the microbottle resonator so that the measured optical readout signal becomes maximum while applying ac voltage from the coherent radio-frequency (rf) source to the electrodes. Note that the optimal horizontal position along the $x$-direction is about 130 µm shifted from the center of the microbottle because the maximum optical mode amplitude appears around 130 µm apart from the center of the microbottle (see Supplemental Materials). At this optimized horizontal position, the gap between the optical and mechanical resonators $d_{\text{OM}}$ is reduced with a nano-positioner. Here it should be noted that the minimum $d_{\text{OM}}$ cannot be zero but finite ($\delta_0$). This is because the optical microbottle physically contacts to the GaAs substrate before contacting the mechanical resonator due to the geometry of the device and the axial profile of the optical microbottle, where $\delta_0$ is estimated to be 250~400 nm from the theoretical approach (see Supplemental Material). In what follows, we alternatively use the parameter $d \equiv d_{\text{OM}} - \delta_0$, which can be experimentally determined such that $d = 0$ where the optomechanical response is saturated.

An external cavity diode laser (ECDL) with the center wavelength of 1030 nm was used to probe the mechanical motion of the beam via the optomechanical coupling. A tapered fiber was used to guide the probe light into a WGM of the optical microbottle resonator. The polarization of the light was properly adjusted to efficiently couple the light into the WGM, and the input power of the light $P_{\text{in}}$ was set to 5.0 µW before the tapered fiber by using appropriate optical components. An avalanche photodiode (APD) was used to detect the light from the tapered fiber. The APD was connected to a bias-T to feed the DC component to a digital sampling oscilloscope (DSO) and the AC component to an electric spectrum analyzer (ESA). In our setup, the laser frequency was scanned to obtain transmission spectra or fixed to the resonance of the microbottle resonator. The former operation was performed with a triangle signal from an arbitrary function generator (AFG) through a servo controller (SC). The transmission spectra in the time domain at the DSO were calibrated by the free spectral range (FSR) of a fiber-loop cavity (FLC)[27]. In the latter operation, a part of the DC component of the transmitted light was sent to the SC as an error signal, and the feedback control was done to stabilize the laser wavelength at the slope of the resonance yielding the highest optomechanical sensitivity.

We observed the optical transmission spectra while reducing the gap $d$, between the optical microbottle resonator and electromechanical resonator (see Fig. 2(a)). The optical Q factor of $1.1 \times 10^6$ was obtained from the full-width at the half maximum (FWHM) at $d = 600$ nm, where the optomechanical coupling is negligible. With decreasing the gap, the amount of the frequency shift $\delta f$ and the linewidth $\kappa$ of the transmission spectra exponentially increased, reflecting the intensity profile of the evanescent field (see Fig. 2(b)). The linewidth became broader by a factor of two while the frequency shift was two times larger than the initial FWHM. The frequency shift and linewidth broadening occurs because the overlap between the optical evanescent field and mechanical beam increases the effective optical cavity length and external dissipation, respectively. In particular, the dissipative nature appeared due to the electromechanical beam thicker than the effective optical wavelength in the beam, which acts as an optical loss channel. Note that the ratio between dispersive and dissipative optomechanical couplings[28] depends on the gap in our system because these exponential factors are different. By driving the mechanical motion with white noise injection, we obtained the mechanical mode spectra via the optomechanical coupling (see Fig. 2(c)). The mode profiles of each spectrum were determined by the finite element method with COMSOL Multiphysics. In the following, we focus on the fundamental mode, whose resonance frequency and mechanical Q factor were $\Omega_\mathrm{M}/2\pi = 282$ kHz and $Q_\mathrm{M} = 1.5 \times 10^3$, respectively. Due to the optical linewidth $\kappa/2\pi = 270$ MHz, our optomechanical system is in an unresolved-sideband regime defined by $\Omega_\mathrm{M}/\kappa \ll 1$, which is suitable for enhancing (heating) and damping (cooling) mechanical motion via feedback control as shown in later. Note that the mechanical resonance frequency is electromechanically controlled by the DC voltages via piezoelectric tension (see Fig.2 (d))[29].

The thermal mechanical motion was measured to quantify the optomechanical coupling. The vacuum optomechanical coupling constant $g_0/2\pi$ was determined by injecting an optical modulation tone from the EOM[30] (see Fig. 3(a)). With decreasing the gap between the microbottle resonator and the beam, $g_0/2\pi$ exponentially increased, reflecting the intensity profiles of the evanescent field, and reached 2.0 Hz. This value is one order of magnitude smaller than the microsphere coupled to a SiN nano-string resonator[31] and three orders of magnitude smaller than on-chip $SiO_2$-SiN optomechanical system[27,32]. Nevertheless, we obtain relatively high displacement sensitivity $S_x^\mathrm{min}$ $\left(\propto g_0^{-1}\kappa\sqrt{P_\mathrm{in}}\right)$ thanks to the high optical Q factor. The displacement sensitivity is determined by $S_x^\mathrm{min} = S_x^\mathrm{th} S_V^\mathrm{back}/S_V^\mathrm{th}(\Omega_M)$, where $S_x^\mathrm{th}$ is the displacement of the thermal mechanical motion at room temperature, and $S_V^\mathrm{back}$ and $S_V^\mathrm{th}(\Omega_M)$ are the measured background noise and the peak of thermal noise, respectively[33]. At $d = 0$ nm ($g_0/2\pi = 2.0$ Hz), we obtain $S_x^{\min} = 3.1 \times 10^2$ fm/$\sqrt{\text{Hz}}$. To improve the displacement sensitivity, the dissipative optomechanical coupling has to be suppressed by decreasing the thickness of the mechanical structure. For instance, purely dispersive optomechanical coupling was achieved with a 30-nm-thick SiN membrane[32]. Using an ultrahigh-Q microbottle resonator[18-23] and a thin electromechanical beam[34], we would be able to perform a displacement measurement near the standard quantum limit ($\sim 0.1$ fm/$\sqrt{\text{Hz}}$ in our system) with the probe power of a few microwatt in our opto-electro-mechanical architecture. Moreover, our optomechanical coupling can be improved by more than one order of magnitude by carefully optimizing the gap with the angled access and the device modification (see Supplemental Material).

The hybridization of the optomechanical system and the electromechanical system allows us to extract the electromechanical conversion coefficient $\eta_{\text{EM}}$ via the optomechanical readout. Here, $\eta_{\text{EM}}$ is a frequency-independent factor with a unit of nm/V, and appears in the electromechanical force as $F_{\text{dr}} = m_{\text{eff}} \Omega_M^2 \eta_{\text{EM}} V_{\text{in}}$ where $m_{\text{eff}}$ is the effective mass, $\Omega_M$ is the angular frequency of mechanical resonance, and $V_{\text{in}}$ is the input voltage. Using the power spectral density (PSD) of the thermal mechanical motion $S_{VV}^{\text{th}}(\Omega)$ and the electromechanical modulation tone $S_{VV}^{\text{Mod}}(\Omega)$ injected by the coherent rf source, $\eta_{\text{EM}}$ is extracted as follows:

$$\eta_{\text{EM}} = \frac{x_{\text{zp}}}{V_{\text{in}}} \sqrt{\frac{2n_M \left\{ \left( \Omega_{\text{Mod}}^2 - \Omega_M^2 \right)^2 + \Gamma_M^2 \Omega_{\text{Mod}}^2 \right\}}{\Omega_M^4}} \sqrt{\frac{\int S_{VV}^{\text{Mod}}(\Omega) d\Omega}{\int S_{VV}^{\text{th}}(\Omega) d\Omega}} \qquad (1)$$

where $x_{\text{zp}}$ is the zero-point fluctuation, $n_M$ is the thermal occupation number, $\Omega_{\text{Mod}}$ is the angular frequency of electromechanical modulation, and $\Gamma_M$ is the linewidth of the thermal mechanical motion (see Supplemental Material). We extracted $\eta_{\text{EM}} = 35$ nm/V, which is reasonably on the order of the reported values in GaAs mechanical resonators[4,24].

In addition to the hybridization of the optomechanical system and the electromechanical system, a feedback loop among optical, mechanical, and electrical systems is implemented by inserting an external electrical circuit after the APD. Here we demonstrate feedback control of the thermomechanical motion using optomechanical detection and piezoelectric feedback driving. This is carried out by inserting a phase adjustor and a low-noise amplifier after APD to enhance or damp the mechanical displacement (see Fig. 4(a)). At $d$=50 nm, we observed both heating and cooling of the mechanical motion by choosing an appropriate feedback phase. The linewidth narrowing (broadening) of the mechanical spectra was clearly observed with the positive (negative) feedback loop (see Figs. 4(b) and (c)). At $d$=0 nm, we performed the further feedback cooling. The effective temperature $T_{\text{eff}}$ defined by the area of the thermal mechanical motion spectra reached 20 K at room temperature. The effective temperature in optomechanical feedback cooling is given by

$$T_{\text{eff}} = \frac{T}{1+g} + \frac{m_{\text{eff}}\Omega_M^2}{4k_B\Gamma_M}\left(\frac{g^2}{1+g}\right)\left(S_x^{\min}\right)^2 \qquad (2)$$

where $T$ is the temperature of the environment, $g$ is feedback gain, and $m_{\text{eff}}$ is the effective mass[35]. Our experimental results show good agreement with the theoretical estimation (see the black line in Fig. 4 (d)). The effective temperature can be further reduced by improving $S_x^{\min}$ with a higher optical Q and smaller dissipative optomechanical coupling.

In conclusion, we developed an opto-electro-mechanical system which possesses both cavity-enhanced optical sensitivity and electrical controllability of the mechanical motion by evanescently coupling a high-Q optical microbottle to a GaAs electromechanical resonator. This evanescent approach allows individual design of optical, mechanical, and electrical systems, which could lead to highly-sensitive and functionalized opto-electro-mechanical systems. Its extension to the microelectromechanical resonators including nitrogen-vacancy centers[36], quantum dot[37], and to the nanoelectromechanical nanowires[38,39], which is not easily integrated with on-chip optical resonators, opens the way towards the construction of hybrid quantum opto-electro-mechanical systems[40].

Supplementary Material

See supplementary material for the theoretical estimation of $g_0$ with respect to the gap and the derivation of Eq. (1).

ACKNOWLEDGEMENT

We acknowledge technical advice from R. Schilling, H. Schütz, V. Sudhir, D. J. Wilson, S. Fedorov, and T. J. Kippenberg, and constructive discussion with Y. M. Blanter. This work is partly supported by MEXT KAKENHI (No. JP15H05869, JP15K21727, JP16H01054, and JP18H04291).

**Reference**

1. M. Aspelmeyer, T. J. Kippenberg, and F. Marquardt, Rev. Mod. Phys. **86**, 1391 (2014).
2. A. N. Cleland, Foundations of Nanomechanics (Springer, New York, 2003).
3. T. Bagci, A. Simonsen, S. Schmid, L. G. Villanueva, E. Zeuthen, J. Appel, J. M. Taylor, A. Sørensen, K. Usami, A. Schliesser, and E. S. Polzik, Nature **507**, 81 (2014).
4. H. Okamoto, A. Gourgout, C-Y. Chang, K. Onomitsu, I. Mahboob, E. Y. Chang, and H. Yamaguchi , Nat. Phys. **9**, 480-484 (2013).
5. T. Faust, J. Rieger, M. J. Seitner, J. P. Kotthaus, and E. M. Weig, Nat. Phys. **9**, 485-488 (2013).
6. I. Mahboob, H. Okamoto, K. Onomitsu, and H. Yamaguchi, Phys. Rev. Lett. **113**, 167203 (2014).
7. D. Hatanaka, T. Darras, I. Mahboob, K. Onomitsu and H. Yamaguchi, Sci. Rep. **7**, 12745 (2017).
8. X-Y. Lü, J-Q. Liao, L. Tian, and F. Nori, Phys. Rev. A **91**, 013834 (2015).
9. M-A. Lemonde, N. Didier, and A. A. Clerk, Nat. Commun. **7**, 11338 (2016).
10. I. W. Frank, P. B. Deotare, M. W. McCutcheon, and M. Lončar, Opt. Express **18**, 8705 (2010).
11. M. Winger, T. D. Blasius, T. P. M. Alegre, A. H. Safavi-Naeini, S. Meenehan, J. Cohen, S. Stobbe, and O. Painter, Opt. Express **19**, 24905 (2011).
12. A. Pitanti, J. M. Fink, A. H. Safavi-Naeini, J. T. Hill, C. U. Lei, A. Tredicucci, and O. Painter, Opt. Express **23**, 3196 (2015).
13. Ž. Zobenica, R. W. van der Heijden, M. Petruzzella, F. Pagliano, R. Leijssen, T. Xia, L. Midolo, M. Cotrufo, Y. Cho, F. W. M. van Otten, E. Verhagen, and A. Fiore, Nat. Commun. **8**, 2216 (2017).
14. K. C. Balram, M. I. Davanço, B. R. Ilic, J-H Kyhm, J. D. Song, and K. Srinivasan, Phys. Rev. Applied **7**, 024008 (2017).
15. C. G. Baker, C. Bekker, D. L. McAuslan, E. Sheridan, and W. P. Bowen, Opt. Express **24**, 20400 (2016).
16. J. Chan, T. P. M. Alegre, A. H. Safavi-Naeini, J. T. Hill, A. Krause, S. Gröblacher, M. Aspelmeyer, and O. Painter, Nature **478**, 89 (2011).
17. L. Ding, C. Baker, P. Senellart, A. Lemaitre, S. Ducci, G. Leo, and I. Favero, Phys. Rev. Lett. **105**, 263903 (2010).
18. J. Volz, M. Scheucher, C. Junge, and A. Rauschenbeutel, Nat. Photon. **8**, 965 (2014)
19. M. Asano, Y. Takeuchi, Ş. K Özdemir, R. Ikuta, L. Yang, N. Imoto, T. Yamamoto, Opt. Express **24**, 12082 (2016).
20. M. Asano, S. Komori, R. Ikuta, N. Imoto, Ş. K Özdemir, T. Yamamoto, Opt. Lett. **41**,

5793 (2016).

21. M. Sumetsky, Phys. Rev. Lett. **111**, 163901 (2013).
22. M. Asano, Y. Takeuchi, W. Chen, Ş. K. Özdemir, R. Ikuta, N. Imoto, L. Yang, T. Yamamoto, Laser & Photon. Rev. **10**, 603 (2016).
23. A. J. R. MacDonald, B. D. Hauer, X. Rojas, P. H. Kim, G. G. Popowich, and J. P. Davis, Phys. Rev. A **93**, 013836 (2016).
24. H. Okamoto, T. Kamada, K. Onomitsu, I. Mahboob, and H. Yamaguchi, Appl. Phys. Express **2**, 062202 (2009).
25. D. Hatanaka, I. Mahboob, H. Okamoto, K. Onomitsu and H. Yamaguchi, Appl. Phys. Lett. **101**, 063102 (2012).
26. R. Ohta, H. Okamoto, and H. Yamaguchi, Appl. Phys. Lett. **110**, 053106 (2017).
27. R. Schilling, H. Schütz, A. H. Ghadimi, V. Sudhir, D. J. Wilson, and T. J. Kippenberg, Phys. Rev. Appl. **5**, 054019 (2016).
28. F. Elste, S. M. Girvin, and A. A. Clerk, Phys. Rev. Lett. **102**, 207209 (2009).
29. A. Kraus, A. Erbe, R. H. Blick, G. Corso, and K. Richter, Appl. Phys. Lett. **79**, 3521 (2001).
30. M. L. Gorodetksy, A. Schliesser, G. Anetsberger, S. Deleglise, and T. J. Kippenberg, Opt. Express **18**, 23236 (2010).
31. G. A. Brawley, M. R. Vanner, P. E. Larsen, S. Schmid, A. Boisen, and W. P. Bowen, Nat. Commun. **7**, 10988 (2016).
32. G. Anetsberger, O. Arcizet, Q. P. Unterreithmeier, R. Rivière, A. Schliesser, E. M.Weig, J. P. Kotthaus, and T. J. Kippenberg, Nat. Phys. **5**, 909 (2009).
33. R. B. Karabalin, M. H. Matheny, X. L. Feng, E. Defaÿ, G. L. Rhun, C. Marcoux, S. Hentz, P. Andreucci, and M. L. Roukes, Appl. Phys. Lett. **95**, 103111 (2009).
34. J. G. E. Harris, D. D. Awschalom, K. D. Maranowski, and A. C. Gossard, Rev. Sci. Instrum. **67**, 3591 (1996).
35. M. Poggio, C. L. Degen, H. J. Mamin, and D. Rugar, Phys. Rev. Lett. **99**, 017201 (2007).
36. S. Kolkowitz, A. C. B. Jayich, Q. P. Unterreithmeier, S. D. Bennett, P. Rabl, J. G. E. Harris, M. D. Lukin, Science **335**, 1603 (2012).
37. Y. Okazaki, I. Mahboob, K. Onomitsu, S. Sasaki, and H. Yamaguchi, Nat. Commun. **7**, 11132 (2016).
38. R. He, X. L. Feng, M. L. Roukes, and P. Yang, Nano Lett. **8**, 1756 (2008).
39. T. S. Abhilash, J. P. Mathew, S. Sengupta, M. R. Gokhale, A. Bhattacharya, and M. M. Deshmukh, Nano Lett. **12**, 6432 (2012).
40. G. Kurizki, P. Bertet, Y. Kubo, K. Mølmer, D. Petrosyan, P. Rabl, and J. Schmiedmayer, Proc. Natl. Acad. Sci. U.S.A. **112**, 3866 (2015).

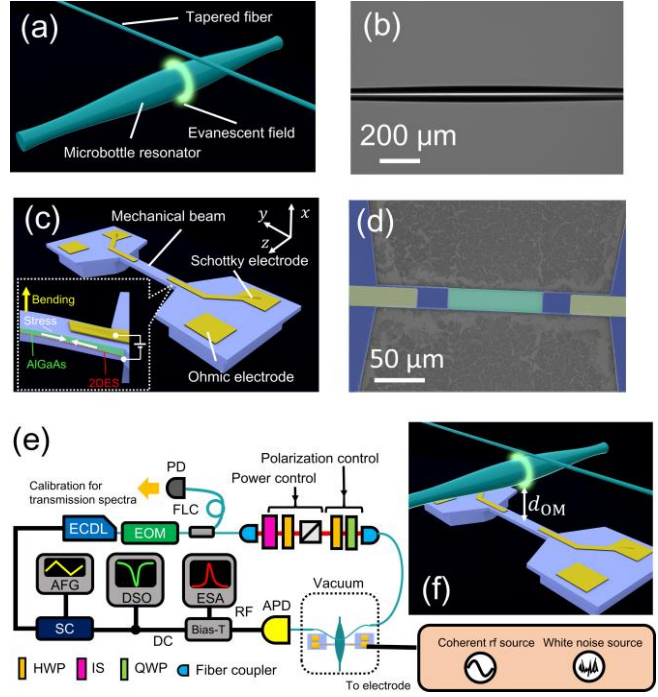

Figure 1: (a) Illustration of an optical microbottle resonator coupled with a tapered fiber. (b) Optical microscope image of the optical microbottle resonator. (c) Illustration of an electromechanical resonator. (d) False color SEM image of the electromechanical resonator based on the AlGaAs/GaAs heterostructure. The blue shaded area is suspended by removing the sacrificial layer, the yellow-shaded area is gold-electrodes, and the green area is a mesa-etched area for electrical isolation. (e) Schematic image of the experimental setup. ECDL: external cavity diode laser. EOM: electro-optic modulator. FLC: fiber-loop cavity. PD: photodiode. APD: avalanche photodiode. ESA: electrical spectrum analyzer. DSO: digital sampling oscilloscope. AFG: arbitrary function generator. SC: servo controller. HWP: half-wave plate. IS: intensity stabilizer. QWP: quarter-wave plate. (f) Illustration of an evanescent optomechanical coupling between the microbottle resonator and electromechanical resonator, which are separated by the gap $d_{\mathrm{OM}}$.

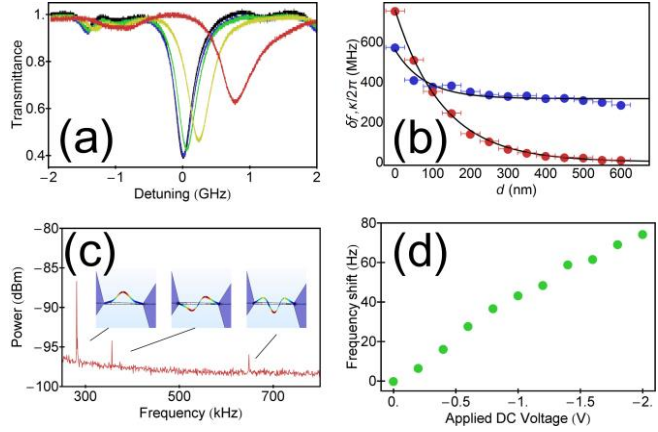

Figure 2: (a) Transmission spectra with respect to the gap between the optical microbottle and electromechanical resonator. The transmission spectra are taken at $d =600$ (black), 450 (blue), 300 (green), 150 (yellow), and 0 nm (red). (b) Frequency shift $\delta f$ (red plots) and linewidth $\kappa/2\pi$ (blue plots) with respect to the gap. The error bar indicates the 50-nm uncertainty coming from the minimum step of the nano-positioner. These are well-fitted by exponential functions shown by the black solid lines. (c) Mechanical mode spectra observed via optomechanical coupling by the excitation of mechanical motions with white noise injection. (d) Frequency shift of the mechanical resonance with respect to the applied DC voltage.

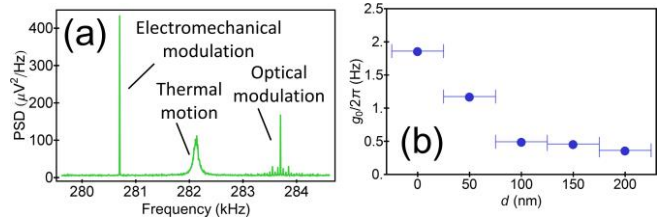

Figure 3: (a) Spectrum of thermal mechanical motion with the two additional modulation tones. (b) Vacuum optomechanical coupling constants with respect to the gap between the optical microbottle and electromechanical resonator. The error bar indicates the 50-nm uncertainty of displacement in the nano-positioner.

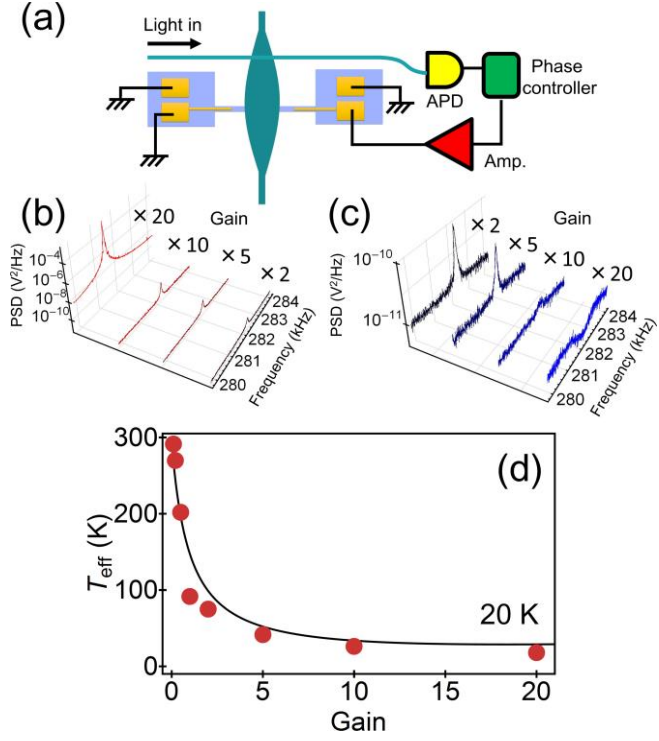

Figure 4: (a) Schematic diagram of the feedback control of mechanical motion with optical detection and electrical control. (b) and (c) Heating and cooling of the fundamental mechanical mode with the positive and negative feedback, respectively. (d) Effective temperatures with respect to the feedback gain. The black line shows the theoretical estimation of effective temperatures with $m_{\text{eff}} = 2.9$ ng, $\Omega_M/2\pi = 282$ kHz, $\Gamma_M/2\pi = 139$ Hz, and $S_x^{\min} = 3.1 \times 10^2$ fm/$\sqrt{\text{Hz}}$.

Supplemental Material: An opto-electro-mechanical system based on evanescently-coupled optical microbottle and electromechanical resonator

1. Theoretical estimation of vacuum optomechanical coupling constant $g_0$ and residual gap $\delta_0$

The vacuum optomechanical coupling constant $g_0$ via the evanescent field from the WGM resonator is given as a function of the gap $d_{OM}$ as follows:

$$g_0(d_{OM}) \approx \omega_0 \alpha \frac{w\sqrt{\frac{\pi R_0}{\alpha}}}{V_{mode}} \frac{1-e^{-2\alpha t}}{2\alpha} (n_{mech}^2 - 1)\xi^2 e^{-2\alpha d_{OM}} \zeta x_{zp}, \quad (S1)$$

where $\alpha^{-1} \equiv \left(\frac{\lambda}{2\pi}\right)\sqrt{n_{opt}^2 - 1}$[S1]. The notation and values used in our calculation is summarized in Tab. S1. Here $\zeta$ is defined as the factor which reflects the mechanical mode profile. For instance, it becomes unity when the optical resonator is placed in the center of mass of the mechanical resonator. In our setup, $g_0$ was measured at the point which is not the center of mass of the beam because the optomechanical coupling is decreased by the mesa structure in which the dimension of the mesa width is comparable to the diameter of the optical microbottle. We experimentally determined $\zeta = 0.7$ by measuring the intensity profile of the fundamental mechanical mode while scanning the position of the optical microbottle (see Fig. S1).

TABLE S1: Parameters for theoretical calculation of $g_0$

| | | |
|---|---|---|
| $\omega_0$ | Angular frequency of optical cavity | $1.88 \times 10^{15}$ Hz |
| $\lambda$ | Wavelength of optical cavity | 1.0 μm |
| $n_{opt}$ | Refractive index of optical cavity (SiO$_2$) | 1.44 |
| $R_0$ | Maximum radius of optical cavity | 40 μm |
| $w$ | Width of mechanical resonator | 20 μm |
| $t$ | Thickness of mechanical resonator | 600 nm |
| $n_{mech}$ | Refractive index of mechanical resonator (GaAs) | 3.49 |
| $\xi$ | Normalized filed intensity at the interface | 0.1 |
| $x_{zp}$ | Displacement at zero-point motion | 3.14 fm |

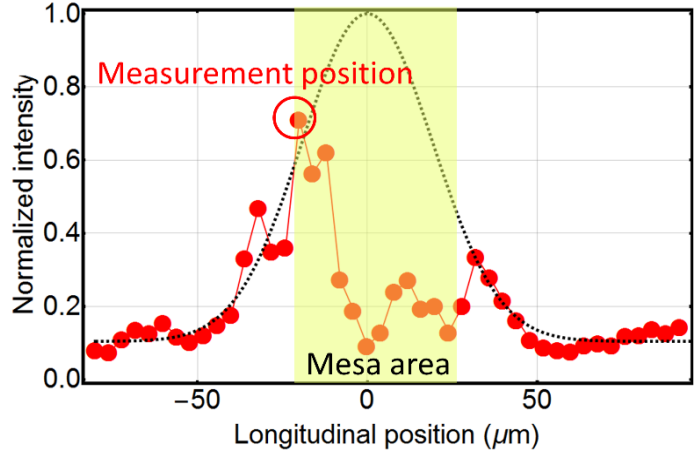

Fig. S1 Intensity profile of the fundamental mechanical mode

The optical mode volume $V_{\text{mode}}$ in the optical microbottle is estimated by assuming that the bottle radius is given as a function of the axial position, $R(z) = R_0\sqrt{1+\frac{1}{2}S^2 z^2}$ where $R_0$ is the maximum radius and $S$ is the curvature of the bottle structure. This assumption allows us to analytically calculate the optical mode $\Psi_{m,q}(r,\phi,z)$ as the product of the radial mode $R_m(r)$ and axial mode $Z_q(z)$ as follows:

$$\Psi_{m,q}(r,\phi,z) = R_m(r)e^{im\phi}Z_q(z) \tag{S2}$$

$$R_m(r) \equiv \begin{cases} J_m(n_{\text{opt}}k_0 r)/J_m(n_{\text{opt}}k_0 R_0) & (r \le R_0) \\ Y_m(k_0 r)/Y_m(k_0 R_0) & (r > R_0) \end{cases} \tag{S3}$$

$$Z_q(z) \equiv H_q\left(\sqrt{\frac{k_\perp S}{2}}z\right)\exp\left[-\frac{k_\perp S}{2\sqrt{2}}z^2\right] \tag{S4}$$

where $(r,\phi,z)$ is the cylindrical coordinate, the $m,q$ are the positive integer which denotes the radial and axial mode indices, $k_0$ is the wavenumber in vacuum, $J_m(\cdot)$ and $Y_m(\cdot)$ denote the first and the second kind of Bessel function, respectively, $S$ is the curvature of the optical microbottle, $H_q(\cdot)$ denotes the Hermite function[S2]. The radial wavenumber $k_\perp$ is determined by the following relationships:

$$k_\perp^2 + k_z^2 = k_0^2 \tag{S5}$$

$$k_z^2 = \frac{2q+1}{\sqrt{2}}Sk_\perp. \tag{S6}$$

The geometry of the optical microbottle and the typical mode profiles are shown in the Fig. S2.

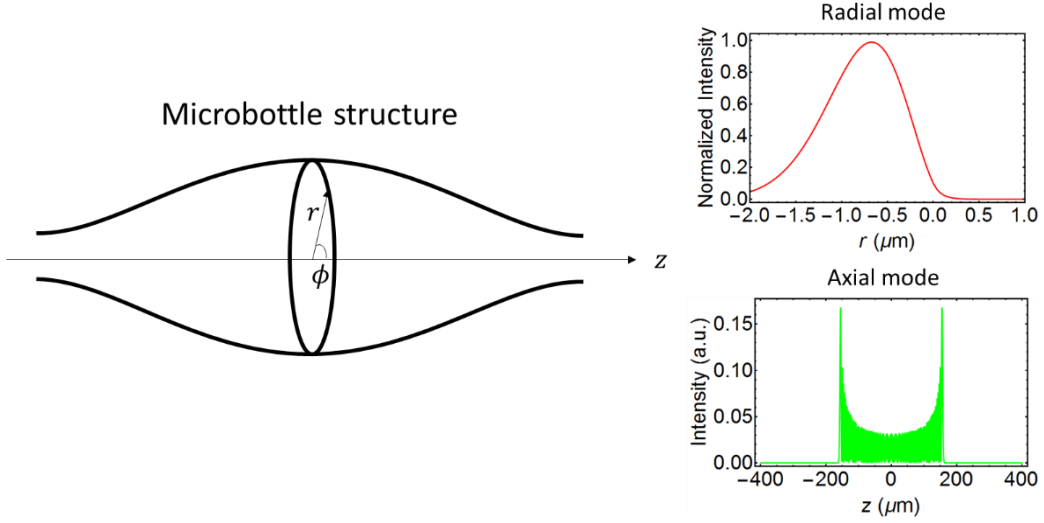

Fig. S2 Geometry of the optical microbottle (left). The typical mode profiles along the radial direction (right top), and the axial direction (right bottom).

The mode volume is approximated to

$$V_{\text{mode}} \approx 2\pi R_0 L_R L_z = 2\pi R_0 \frac{\int n^2(r)|R_m(r)|^2 dr}{\max_r[n^2(r)|R_m(r)|^2]} \frac{\int n^2(z)|Z_q(r)|^2 dz}{\max_z\left[n^2(z)|Z_q(z)|^2\right]}. \quad (S7)$$

Because it is not easy to exactly determine the mode indices experimentally, we estimate $V_{\text{mode}}$ with a finite margin which is reasonable in our experimental setup. First, we determine the radial mode length $L_R \equiv \frac{\int n^2(r)|R_m(r)|^2 dr}{\max_r[n^2(r)|R_m(r)|^2]}$. The minimum radial mode length is obtained $L_R^{\min} = 1.4$ μm by setting $m = 350$ which corresponds to the fundamental radial optical modes at the wavelength of 1 μm. Because the tapered fiber was physically contacted to the optical microbottle in our experiment, the radial mode may not be the fundamental mode. Thus, we also take into account of the 7th-order radial mode which has a twice of the radial mode length $L_R^{\max} = 2.8 \times 10^{-6}$ m (see Fig. S3). Secondly, we determine the axial mode length $L_z = \frac{\int n^2(z)|Z_q(r)|^2 dz}{\max_z\left[n^2(z)|Z_q(z)|^2\right]}$. We roughly estimate the axial mode indices of the optical microbottle from the optical microscope image, which shows the position of the optical microbottle associated with the mechanical beam (see Fig. S4). We assume that the yellow shaded area corresponds to the center (z~0), and obtain the distance from the center to the point crossing to the mechanical beam. Because the optomechanical coupling was optimized with respect to the axial position of the optical microbottle, we estimate the two axial mode profiles, which indicate the intensity peak at the distance of $z_c = 130 \pm 50$ μm from the center. These distances correspond to the axial mode numbers q=50 and q=200 whose axial mode length are $L_Z^{min} = 53\ \mu m$ and $L_Z^{max} = 67\ \mu m$, respectively. Finally, $V_{mode}$ is estimated with the finite margin from $V_{mode}^{min} = 1.9 \times 10^4\ \mu^3 m$ to $V_{mode}^{max} = 4.9 \times 10^4\ \mu^3 m$.

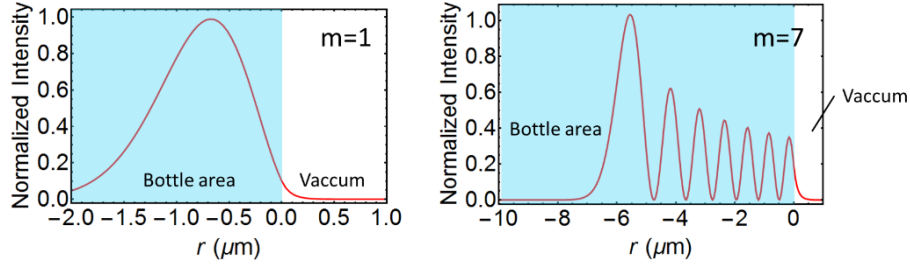

Fig.S3 Radial intensity profile of the fundamental and 7$^{th}$ radial mode.

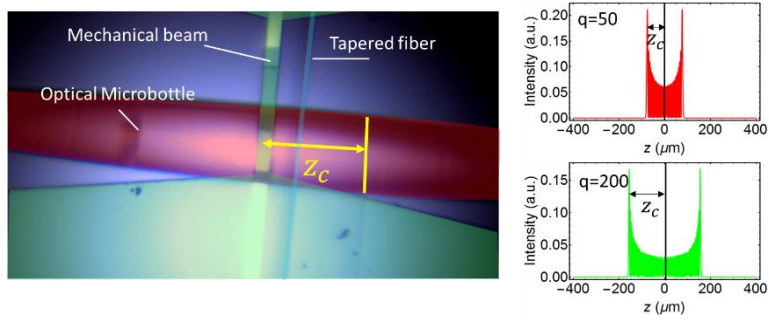

Fig. S4 Optical microscope image in our experimental setup (left). The green shaded area is the mechanical beam structure, the blue shaded area is the optical tapered fiber, and the red shaded area is the optical microbottle. The yellow shaded area corresponds to the estimated center of the optical microbottle. The optical axial mode profiles which indicate the peak intensity at the point with the distance $z_c = 80\ \mu m$ (red) and $180\ \mu m$ (green) from the center (right).

By instituting the estimated mode volumes into Eq. S1, we obtain expected $g_0/2\pi$ as the function of the gap $d_{OM}$ (see Fig. S5).

We could not obtain $d_{OM} = 0$ experimentally, because the difference between the radius at the center and one at the point crossing to the beam is comparable to the depth of the sacrificial layer of the beam structure (see Fig. S6). The residual gap $\delta_0$ is roughly estimated to 250~400 nm from the correspondence between the theoretical estimation and the experimental values of $g_0$. These estimation implies that we can expect $g_0$ which is two order of magnitude larger than the obtained value by increasing the depth of the sacrificial layer or by selecting the optical mode which localized around the center of the optical microbottle in order to obtain $\delta_0 = 0$.

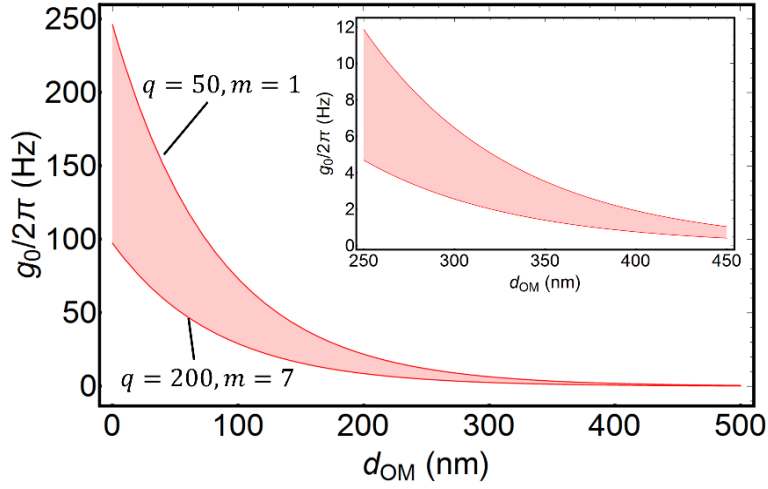

Fig. S5 Estimated vacuum optomechanical coupling constant $g_0/2\pi$ with respect to the gap $d_{OM}$. The inset shows the magnified area around the residual gap $\delta_0$.

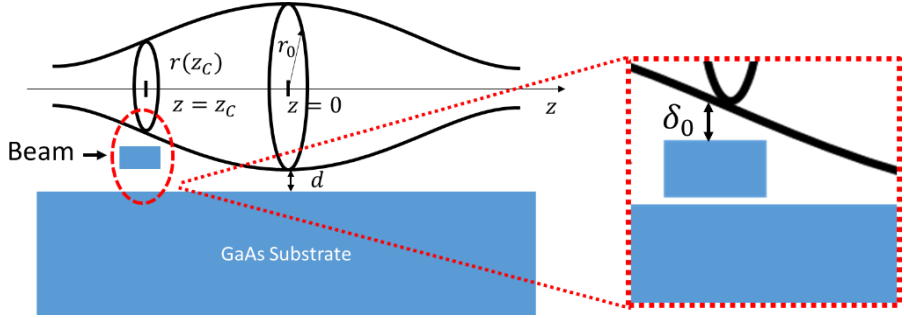

S6. Residual gap due to the geometry of the optical microbottle. $d \equiv d_{OM} - \delta_0$ is the gap experimentally determined.

## 2. Derivation of electromechanical conversion coefficient

The displacement of mechanical motion $x(t)$ obeys the equation

$$m_{\text{eff}}\ddot{x}(t) - m_{\text{eff}}\Gamma_M \dot{x}(t) + m_{\text{eff}}\Omega_M^2 x(t) = F_{\text{EM}} \tag{S8}$$

where $m_{\text{eff}}$, $\Gamma_M$, $\Omega_M$ are the effective mass, the intrinsic damping constant, and the angular frequency of the mechanical mode. We suppose that the external force induced by electromechanical conversion coefficient linearly depends on the input voltage $V_{\text{in}}$ as below:

$$F_{\text{EM}} \equiv m_{\text{eff}}\Omega_M^2 \eta_{\text{EM}} V_{\text{in}} e^{-i\Omega_{\text{Mod}} t} \tag{S9}$$

where $\Omega_{\text{Mod}}$ is the angular frequency of the input voltage, and $\eta_{EM}$ is the electromechanical conversion coefficient. In the frequency domain, the linear susceptibility $\chi_M(\Omega) \equiv m_{\text{eff}}(\Omega^2 - \Omega_M^2 + i\Gamma_M \Omega)$ formulates the displacement in frequency domain

$$\tilde{x}(\Omega) = \chi_\mathrm{M}^{-1}(\Omega) m_\mathrm{eff}\Omega_\mathrm{M}^2 \eta_\mathrm{EM} V_\mathrm{in}\delta(\Omega - \Omega_\mathrm{Mod}). \tag{S10}$$

Then, the integral of power spectrum density (PSD) of the external force is given by

$$\int S_{xx}^\mathrm{Mod}(\Omega)\mathrm{d}\Omega = |\chi^{-1}(\Omega_\mathrm{Mod})|^2 m_\mathrm{eff}^2 \Omega_\mathrm{M}^4 \eta_\mathrm{EM}^2 V_\mathrm{in}^2. \tag{S11}$$

At the thermal equilibrium, the integral of PSD of thermal mechanical motion is formulated by the zero-point fluctuation $x_\mathrm{zp}$ and the thermal occupation number $n_\mathrm{M}^\mathrm{th}$ as below:

$$\int S_{xx}^\mathrm{th}(\Omega)\mathrm{d}\Omega = 2x_\mathrm{zp}^2 n_\mathrm{M}^\mathrm{th}. \tag{S12}$$

Since the PSD of thermal mechanical motion is fully characterized by the dimension of the device and temperature, it is able to use as a reference to determine $\eta_\mathrm{EM}$. We do not directly obtain the PSDs of displacement, just obtain the PSDs of electrical voltage from a photodetector in the practical experiment. In order to make a correspondence between them, we suppose that the ratio among the PSDs of electrical voltage are the same as the ones of displacement as below:

$$\sqrt{\frac{\int S_{VV}^\mathrm{Mod}(\Omega)d\Omega}{\int S_{VV}^\mathrm{th}(\Omega)d\Omega}} = \sqrt{\frac{\int S_{xx}^\mathrm{Mod}(\Omega)d\Omega}{\int S_{xx}^\mathrm{th}(\Omega)d\Omega}}. \tag{S13}$$

Eq. (S13) has been often used to determine the vacuum optomechanical coupling[S3]. By substituting Eq. (S11) and (S12) into (S13), we obtain

$$\sqrt{\frac{\int S_{VV}^\mathrm{Mod}(\Omega)d\Omega}{\int S_{VV}^\mathrm{th}(\Omega)d\Omega}} = \sqrt{\frac{2x_{zp}^2 n_M^\mathrm{th}}{\eta_{EM}^2 m_\mathrm{eff}^2 \Omega_\mathrm{M}^4 |\chi_M(\Omega_\mathrm{Mod})|^2 V_\mathrm{in}^2}}. \tag{S14}$$

Then, the electromechanical conversion coefficient is expressed by

$$\begin{aligned}\eta_{EM} &= \sqrt{\frac{2x_{zp}^2 n_M^\mathrm{th}}{m_\mathrm{eff}^2 \Omega_\mathrm{M}^4 |\chi_M(\Omega_\mathrm{Mod})|^2 V_\mathrm{in}^2}} \sqrt{\frac{\int S_{VV}^\mathrm{th}(\Omega)d\Omega}{\int S_{VV}^\mathrm{Mod}(\Omega)d\Omega}} \\ &= \frac{x_{zp}}{V_\mathrm{in}}\sqrt{\frac{2\{(\Omega_\mathrm{Mod} - \Omega_\mathrm{M})^2 + \Gamma_\mathrm{M}^2 \Omega_\mathrm{Mod}^2\}}{\Omega_\mathrm{M}^4}} \sqrt{\frac{\int S_{VV}^\mathrm{th}(\Omega)d\Omega}{\int S_{VV}^\mathrm{Mod}(\Omega)d\Omega}}.\end{aligned} \tag{S15}$$

S1. G. Anetsberger, O. Arcizet, Q. P. Unterreithmeier, R. Rivière, A. Schliesser, E. M.Weig, J. P. Kotthaus, and T. J. Kippenberg, Nat. Phys. **5**, 909 (2009).

S2. Y. Louyer, D. Meschede, and A. Rauschenbeutel, Phys. Rev. A **72**, 031801 (2005).

S3. M. L. Gorodetksy, A. Schliesser, G. Anetsberger, S. Deleglise, and T. J. Kippenberg, Opt. Express **18**, 23236 (2010).